# Few-Shot Classification of Autism Spectrum Disorder using Site-Agnostic Meta-Learning and Brain MRI

Nikhil J. Dhinagar*, *Member, IEEE,* Vignesh Santhalingam,
Katherine E. Lawrence, Emily Laltoo, Paul M. Thompson

*Abstract*— For machine learning applications in medical imaging, the availability of training data is often limited, which hampers the design of radiological classifiers for subtle conditions such as autism spectrum disorder (ASD). Transfer learning is one method to counter this problem of low training data regimes. Here we explore the use of meta-learning for very low data regimes in the context of having prior data from multiple sites – an approach we term site-agnostic meta-learning. Inspired by the effectiveness of meta-learning for optimizing a model across multiple tasks, here we propose a framework to adapt it to learn across multiple sites. We tested our meta-learning model for classifying ASD versus typically developing controls in 2,201 T1-weighted (T1-w) MRI scans collected from 38 imaging sites as part of Autism Brain Imaging Data Exchange (ABIDE) [age: 5.2 -64.0 years]. The method was trained to find a good initialization state for our model that can quickly adapt to data from new unseen sites by fine-tuning on the limited data that is available. The proposed method achieved an ROC-AUC=0.857 on 370 scans from 7 unseen sites in ABIDE using a few-shot setting of 2-way 20-shot i.e., 20 training samples per site. Our results outperformed a transfer learning baseline by generalizing across a wider range of sites as well as other related prior work. We also tested our model in a zero-shot setting on an independent test site without any additional fine-tuning. Our experiments show the promise of the proposed site-agnostic meta-learning framework for challenging neuroimaging tasks involving multi-site heterogeneity with limited availability of training data.

*Clinical Relevance*— We propose a learning framework that accommodates multi-site heterogeneity and limited training data to assist in challenging neuroimaging tasks.

## I. INTRODUCTION

Autism spectrum disorder (ASD) is a neurodevelopmental disorder that may be detected in early childhood, but often may not be diagnosed until later during adult life. According to the World Health Organization (WHO), about 1 percent of children worldwide have autism [1].

In this work, we were interested in tackling MRI-based ASD classification using data from multiple sites in a meta-learning framework. Large-scale MRI studies of ASD [2] have found robust differences in brain morphometry in ASD but with low effect sizes. In a case-control mega-analysis of 1,571 individuals with ASD and 1,651 healthy controls (aged 2-64 years) from 49 participating sites, ASD was associated with smaller subcortical volumes of the pallidum, putamen, amygdala, and nucleus accumbens (effect sizes [Cohen's *d*]: 0.13 to -0.13), as well as higher cortical thickness in the frontal cortex and lower thickness in the temporal cortex (effect sizes: -0.21 to 0.20). Some studies have reported ASD classification accuracy around 70% when using deep learning and resting-state functional MRI data [3].

There is increasing interest in applying machine learning and deep learning methods to the more standard type of MRI scan used radiologically - T1-weighted (T1-w) anatomical MRI. A recent review notes that this task is promising but challenging [4]. Traut et al. [5] summarize insights from an ASD imaging biomarker challenge. They show that a prediction accuracy of AUC 0.66 was obtained using only anatomical MRI. Haar et al., [6] obtained an accuracy between 56-60% accuracy using T1-w data from the ABIDE dataset. Katuwal et al., achieved an AUC of 0.61 using structural MRIs [7]. Gao [8] was able to obtain an AUC of 0.67 by creating individual-level morphological covariance brain networks constructed from the structural MRIs from ABIDE. Sapanaro et al. [9] obtained an AUC of 0.58, using only male subjects from the ABIDE dataset.

Prior works attribute the slightly above chance-level results in ASD classification from the large-scale multi-ste ABIDE dataset to ASD heterogeneity [9] [6] [7] [14]. In this work we propose an approach we term a site-agnostic meta-learning model to create an ASD classifier that is robust to site-based diversity and limited training data. Finn et al. originally introduced Model-agnostic Meta-learning (MAML) [10] for rapidly adapting deep neural networks to new tasks. This method was successful in learning an optimal parameter space by traversing task-specific parameters and backpropagating over the aggregated losses on a separate dataset to update a meta model. Several variants of MAML have since been published to enhance the original MAML algorithm, including first-order MAML or *reptile* [11], iMAML [12], MAML++ [13], and almost no inner loop MAML [14]. In MAML++, the authors propose several ideas to update MAML, including addition of a multi-step loss. Ouahab et al. proposed a meta-learning approach for few-shot disease classification [15].

In our work, the meta-model is trained over multiple random combinations of the sites in a few-shot paradigm of 2-way, 20-shot setting from the ABIDE dataset to obtain an optimal initialization. Here, 20-shot indicates the number of samples per site. The meta-model can then rapidly adapt to data from new unseen sites with minimal fine-tuning during inference. Our main contributions are that we:

1. Propose a meta-learning framework for ASD classification, that is robust to site-based heterogeneity in structural T1-w MRI scans,
2. Test how well the proposed meta-model can adapt to new unseen sites with limited data in a few-shot setting, and
3. Zero-shot test the model on data from an independent test site without any additional fine-tuning to evaluate model generalization.

## II. DATA

We analyzed 2,201 T1-w brain MRI scans collected from 38 imaging sites as part of the Autism Brain Imaging Data Exchange (ABIDE). Fig. 1 captures the heterogeneity of the ABIDE dataset in terms of age, sex, and diagnosis. The data was prepared carefully for testing the meta-learning methods as well as on baseline methods. 30 (i.e., 38-7-1) sites were used as part of the meta-training phase as shown in Table 1.

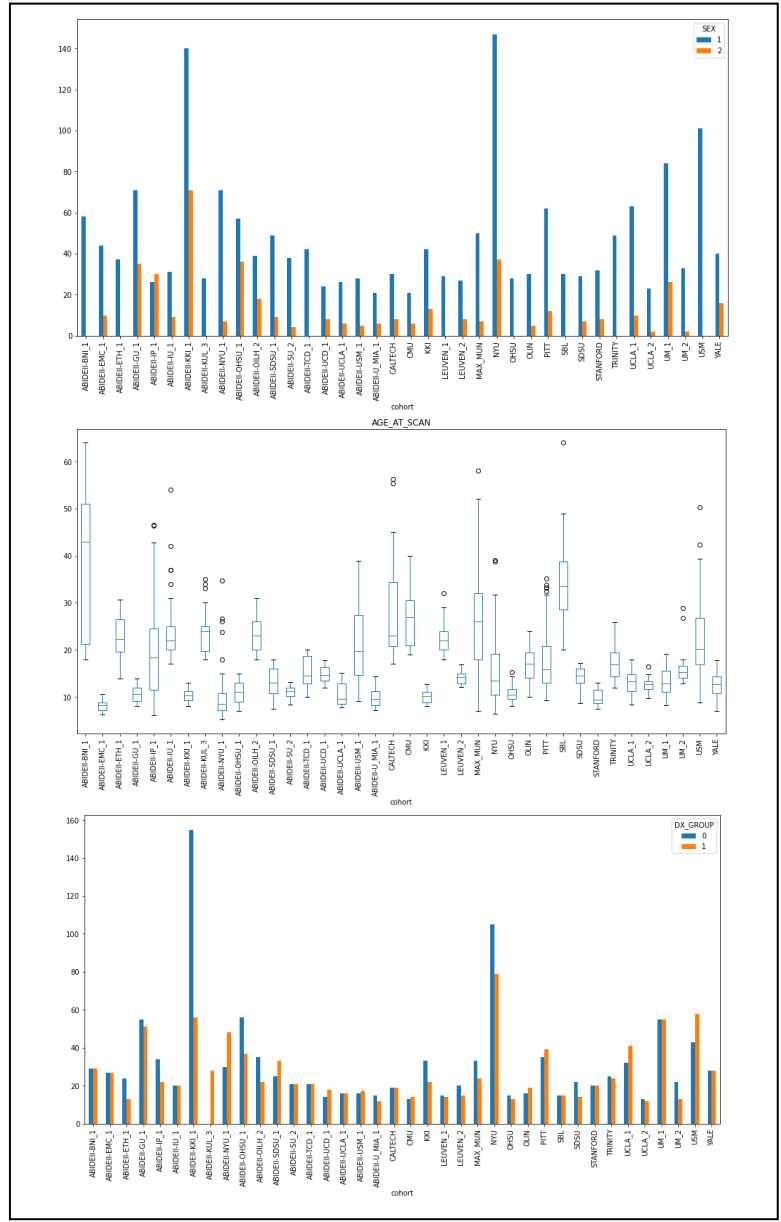

Figure 1. Heterogeneity in the ABIDE dataset captured across the 38 sites via plots of key clinical variables. X-axis - site, Y-axis - number of samples for the variable of interest (by sex/age/diagnosis). Top - bar plot of the sex (Male '1'/Female '2') distribution, box plot of the age distribution, and bar plot of the diagnosis (control '0'/ASD '1').

TABLE I. DATA USED FOR THE META-LEARNING FRAMEWORK.

|  | Meta-training: Train | Meta-training: Validation | Meta-testing | Zero-shot Test |
|---|---|---|---|---|
| N | 1317 | 330 | 370 | 184 |
| Age Range in Years (average[SD]) | 5.2-64.0 (15.23 [7.68]) | 5.3-42.8 (14.76 [6.78]) | 7.0-64.0 (20.59 [12.23]) | 6.5-39.1 (15.25 [6.56]) |
| Females | 280 (21.3%) | 57 (17.3%) | 47 (12.7%) | 37 (20.1%) |
| Autism Patients | 614 (46.6%) | 161 (48.8%) | 175 (47.3%) | 79 (42.9%) |
| #Sites | 30 | 30 | 7 | 1 |

Further, we used data from 7 sites for meta-testing and from one site as the out-of-distribution zero-shot test set. For our baseline, we used three sites from the meta-test dataset to simulate the few-shot setting. Here the data was divided such that the training set had 20 samples per site and the rest of the samples were divided into validation/testing.

We used raw T1-w scans from each of the ABIDE sites without any prior pre-processing. The input images for all the models were z-scored based on the mean and standard deviation, to standardize them. For the meta-learning model, the scans were prepared by first resizing the images via interpolation to a standard spatial dimension of 91x91x91. Then the slices from each 3D brain scan were arranged in three rows on the spatial plane, where each row corresponds to one of the orientations, i.e., axial, coronal and sagittal. Lastly, every fifth slice was then selected, and the resultant image was down sampled by 0.25, to further reduce system memory requirements during model training.

## III. METHODS

### A. Baseline Model

We used a 3D DenseNet121[16] CNN, commonly used for different neuroimaging applications [17]. We also used a scaled-down version of the DenseNet121 called the TinyDenseNet, with fewer parameters [18].

Further, we used transfer learning to create an initialization for our model. We pre-trained our model on 30 ABIDE sites in a supervised manner for ASD classification. As part of our baseline, we also fine-tuned our ABIDE pre-trained model on our ABIDE meta-test data set with 20 samples per site from 3 sites.

### B. Proposed Method: Site-Agnostic Meta-Learning

Model agnostic meta-learning (MAML) [10] is an optimization-based meta-learning approach to find an optimal set of model weights that can be quickly adapted to a new task in a few gradient steps. Finn et al. [10] express meta-learning as a superset of the 'fine-tuning' paradigm, by testing the model 'fine-tuned' on a support set, on a separate (target) dataset across multiple tasks, which is then used to optimize overall model parameters as shown in equation 1.

$$\min_\theta \sum_{site\ i} \mathcal{L}\ (\theta - \alpha \nabla_\theta \mathcal{L}\left(f_\theta D_i^S\right), D_i^T) \quad (1)$$

where $\theta$ are the model parameters, $\mathcal{L}$ is the loss function, $D_i^S$ is the support set, $D_i^T$ is the target set, $\alpha$ is the learning rate.

The general recipe for MAML and meta-learning involves meta-training and meta-testing. The original MAML approach, though powerful, can be computationally expensive; one of the reasons is computation of the target loss after all the inner loop updates with each support set. Here we used the MAML++ [13] framework, which includes certain modifications to stabilize training. such as calculating the target set loss after every step on the support set. Typically, MAML and meta-learning are used to learn across multiple tasks. In this work, we formulate the meta-learning framework to learn across data from multiple sites, for our ASD classification task.

The backbone used in our work is based on the VGG architecture [19]. Meta-training produces a 'meta-model' similar to a pre-trained model that is later 'fine-tuned' and tested during the following meta-testing stage as in Fig. 2.

### B.1 Meta-Training

The meta-training algorithm that we propose is described step-by-step in Fig. 3. First, each site's data is randomly sampled into a support set and disjoint target set from each of the sites during meta-training. The number of sites, support set sample size per site (i.e., N-site, K-samples per site) and target set sample size per site are treated as hyperparameters that were selected based on a random search to maximize the validation AUC-ROC from a range of values (number of sites $\varepsilon$ {1, 2, 3, 4, 5, 6} and the number of support samples per site $\varepsilon$ {10, 12, 15, 18, 20} and the target samples/site $\varepsilon$ {2, 5, 8, 10}). We selected the optimal values to be 1-site, 20-support samples per site, and 10 target samples per site.

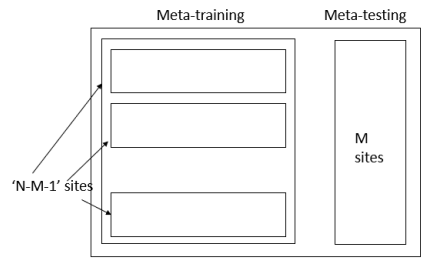

Figure 2. The proposed framework for a Meta-Learning approach learns an optimal parameter space for fast adaptation to previously unseen sites (M) with limited training data.

```
Proposed Algorithm. Meta-Training: Learning from multiple sites
─────────────────────────────────────────────────────────────
Require: p(S): distribution over sites ε D_meta-train
Require: α, β, learning and meta-learning rates
1: randomly initialize θ
2. while number of epochs or early stopping is not done do  //outer loop
3.    Sample batch of sites S_i ~ p(S) ε D_meta-train
4.    while inner loop steps is not done for each site S_i do // inner loop
5.       Sample support set D_i^s = {x_i, y_i} from S_i
6.       forward pass with D_i^s and obtain gradients
7.       compute updated parameters for inner loop model
            θ'_i ← θ − α∇_θ L_{Si}(f_θ D_i^s)
8.       Sample target set D_i^t = {x_i, y_i} from S_i
9.       forward pass with D_i^t and accumulate loss for meta-update
10.   end while
11.   Compute metrics for D_i^t predictions
12.   update the meta-model parameters θ
13. end while
```

Figure 3. Algorithm description for our proposed site-agnostic meta-learning framework for ASD classification.

Using this data, meta-training was carried out with two main stages, the inner loop and the outer loop update. The outer loop maintains an overall 'meta-model' that is updated based on each site-specific model update in the inner loop. During each iteration of the inner loop, a copy of the model is trained to classify ASD on a randomly sampled support set of 20 scans from a single site out of total of 30 sites (i.e., 2-way, 20-shot). This 'fine-tuned' model is then tested on a target set of 10 samples from each respective site. The loss aggregated from each of the forward passes on the target set from each site is then used to update the meta model's parameters. Following each meta-training epoch, the model is evaluated on a validation set, where the top 5 models are recorded over training time.

### B.2 Meta-Testing

During meta-testing, the top 5 meta-models from the meta-training phase were fine-tuned and tested on data from previously unseen sites. The fine-tuning occurs within the few-shot learning paradigm of 2-way, 20-shot - i.e., 20 samples per site. The meta-model was 'fine-tuned' on 20 support samples per site and tested on a disjoint 10 samples from the same site without replacement. The final test performance of the top model was aggregated over all the 370 scans from the meta-test dataset from the 7 unseen sites.

### C. Hyperparameter Optimization and Model Training Strategy

For the baseline CNNs, we used a batch size of 16, a cosine learning rate scheduler, Adam optimizer with weight decay of 1e-4, and early stopping at 20 epochs. We used a learning rate of 3e-4 when training from scratch, and 1e-4 when fine-tuning a pre-trained model.

For the meta-learning framework, we used a batch size of 1, a cosine learning rate schedule, Adam optimizer with weight decay of 1e-4 for site-agnostic learning and Stochastic-Gradient-Descent-Learning-Rule Optimizer for tuning the per-layer learnable learning rates. We minimized a binary cross-entropy loss function for all our models.

## IV. RESULTS

Our proposed method, site-agnostic meta-learning, achieved a 1-site, 20-shot performance with an ROC-AUC of 0.850 and a balanced accuracy of 0.80 (weighted by the samples per class) as shown in Table II. The proposed method also obtained a ROC-AUC of 0.622 on a single site test set in a zero-shot manner. The proposed method also obtained a ROC-AUC of 0.622 on a single site test set in a zero-shot manner. Our best baseline - with the TinyDenseNet CNN pre-trained on data from 30 sites in ABIDE - achieved a test ROC-AUC of 0.613 when trained on 20 images per site from 3 unseen sites. This baseline achieved a ROC-AUC of 0.558 on the zero-shot test set.

TABLE II. PERFORMANCE OF THE PROPOSED SITE-AGNOSTIC META-LEARNING METHOD, FOR A 2-WAY, 20-SHOT ASD CLASSIFICATION (20 TRAINING SCANS/SITE) ON A TEST SET OF 370 SCANS ACROSS 7-SITES - IN COMPARISON WITH BASELINE CNNS AND TRANSFER LEARNING. . THE BASELINE METHODS - USING THE DENSENET AND TINYDENSENET ARCHITECTURES - WERE TESTED ON A SUBSET OF THE META-TEST SET WITH THREE SITES (TRAIN: 72 SCANS (24 SCANS/SITE), VALIDATION: 18 SCANS (6 SCANS/SITE), TEST: 135 SCANS ACROSS THREE SITES) TO SIMULATE THE FEW-SHOT LEARNING PARADIGM. RESULTS SHOW THAT THE PROPOSED MODEL GENERALIZES BETTER TO A WIDER RANGE OF SITES FROM THE SAME DATASET.

| Method/ Architecture | Pre-training | Meta-Test Test ROC-AUC | Zero-shot single-site Test ROC-AUC (without fine-tuning) |
|---|---|---|---|
| Site-Agnostic Meta-Learning (Proposed) | - | **0.850** | 0.622 |
| TinyDenseNet CNN (transfer learning) | ABIDE 30 sites | 0.613 | 0.558 |
| TinyDenseNet CNN | - | 0.542 | 0.542 |
| DenseNet121 CNN | - | 0.530 | 0.386 |

## V. Discussion

Typically, in limited data settings, transfer learning is the preferred solution to fine-tune large pre-trained models for a downstream task. In our experiments, our proposed method based on meta-learning outperformed our transfer learning baseline in a few-shot setting, with a 24% improvement in the ROC-AUC. Our method achieved this performance when fine-tuned on only 20 samples per site. Our approach also outperformed the baseline by a 6% increase in the ROC-AUC given an independent single site dataset in zero-shot setting, i.e., without requiring any additional fine-tuning.

Prior works report an AUC around 0.66-0.67 for ASD classification based on structural MRIs. We attribute the significant improvement in performance as we explicitly designed our system to learn site-based differences to address heterogeneity in the ABIDE dataset. The formulation of the gradient-based MAML approach for few-shot learning lent itself to our goal of adapting a model across multiple sites. A similar improvement in performance using a MAML-like approach was also observed in prior work by Ouahab et al., [7] in applying few shot classification for skin lesion classification. In our experiments, we found that it was optimal to use a single site during each of the meta-model updates. This is intuitive as the model can then be 'fine-tuned' on data from each of the sites over time; this can overcome the heterogeneity in the dataset across all of the sites following meta-training.

Katuwal et al. [7] report higher performance using subgroup analysis on the ABIDE dataset. In future work, we plan to further evaluate our proposed method using a cross-validation approach such that a diverse number of sites are seen during both the meta-training and meta-testing phases.

## VI. Conclusion

In this paper, we have proposed a method that we term, site-agnostic meta-learning, a framework for ASD classification using structural brain MRI scans. We show the potential to accommodate multi-site heterogeneity using our meta-learning approach. Our experiments show that this method used in a few-shot learning context could be useful in neuroimaging applications with limited data availability.

## Acknowledgment

This research was supported by the NIMH award F32MH122057 and Momental Foundation Mistletoe Research Fellowship to K.E.L., and U01AG068057.